\documentclass[pre, %
reprint,
 amsmath,amssymb,
 aps,
floatfix,
]{revtex4-2}

\usepackage[hidelinks]{hyperref}
\usepackage{graphicx}
\usepackage{dcolumn}
\usepackage{bm}
\usepackage{layout}
\usepackage{float}

\begin{document}

\preprint{APS/123-QED}

\title{Diffusion-driven pattern formation in an opinion dynamical network model}

\author{Tim Mauch}
 \email{tim.mauch1@gmail.com}
\author{Thilo Gross}%
 \affiliation{Helmholtz Institute for Functional Marine Biodiversity (HIFMB), Im Technologiepark 5, 26129 Oldenburg, Germany}
 \affiliation{Carl-von-Ossietzky University, Institute for Chemistry and Biology of the Marine Environment, Carl-von-Ossietzky Str. 9-11, 26129 Oldenburg, Germany}
\affiliation{Alfred-Wegener Institute (AWI), Helmholtz Center for Polar and Marine Research, Am Handelshafen 12, 27570 Bremerhaven, Germany}

\date{\today}

\begin{abstract}
The spatial organization of individuals and their interactions in communities are important factors known to preserve diversity in many complex systems. Inspired by metapopulation models from ecology, we study opinion formation using a network-based approach in which nodes represent communities of interacting agents holding one of two competing opinions, and links represent avenues of migration. Agents adapt to the dominant opinion within a community or migrate toward other communities.
Using a master stability function approach, we analytically derive conditions for diffusion-driven pattern formation and identify structural features of the community network that sustain opinion diversity. Our model shows that even under minimal opinion rules, the interaction between local dynamics and community structure generates spatial patterns that allow minority opinions to persist by gaining local dominance.
\end{abstract}


\maketitle


\section{\label{sec:introduction}Introduction}
Consider a system where individuals can migrate in space while simultaneously being reshaped by their environment. Schelling introduced a seminal model in which he captured this logic, where individuals relocate when their environment becomes too dissimilar, thereby demonstrating how simple local preference can lead to spatial segregation \cite{Schelling1971}. Examples of such systems are plentiful: in U.S. politics, it is now common that people prefer to live in communities that are perceived to align with their political opinions, while simultaneously those opinions are being reshaped by the communities. Similarly, in the online space, we experienced a recent mass exodus from the platform formerly known as Twitter, following a perceived ideological shift.

The theme of environment reshaping individuals while individuals move in response to the environment is not limited to humans and human opinions. In ecology, behavioral patterns of animals are transmitted very similarly to human opinions \cite{Axelrod1981, BoydRicherson1985}. Moreover, comparable dynamics also exist on the genetic level, where mortality and reproduction can lead to a competition between alleles of a gene that are modeled in the same way as competition between opinions \cite{Lieberman2005}. For both genetic and behavioral evolution of animals, dispersal of individuals between habitats is known to play a significant role for the creation and maintenance of diversity \cite{Lieberman2005, NowakMay1992, Fahimipour2022, Webber2023}.

In science, the study of human and animal behavior have long profited from network models, leading to a history of cross-fertilization between models \cite{Lieberman2005, Traulsen2010}. Arguably, the most basic model of opinion formation is the voter model, which was proposed by Holley and Liggett \cite{Holley1975}, but was preceded by a mathematically identical model for the replacement of species that was originally proposed by Clifford and Sudbury \cite{Clifford1973}.

Refinements to the opinion formation models included continuous-valued opinion formation, where opinions are modeled as points on an axis, representing, for example, the political spectrum \cite{Deffuant2000, Hagselmann2002}. For discrete opinions, more detailed modeling of the opinion adoption process led to complex-contagion models, in which the rate at which agents adopt new opinions depends nonlinearly on the agents' neighborhood \cite{Granovetter1978, Watts2002}. Finally, adaptive networks were proposed to model situations in which agents can actively reshape their neighborhood by cutting or rewiring connections \cite{Gross_2007, HolmeNewman2006, Min_2017}. Particularly, the active reshaping of connections was also considered for opinion formation in groups of animals, where the opinions under consideration concern the movement of the group \cite{Couzin_2011, Chen2016, Fahimipour2022}.

The predictions of opinion formation models have been tested in computerized experiments with humans, collective motion experiments with animals and by analysis of social media and election data \cite{Gracia2014, Couzin_2011, Threlkeld2017, Vendeville2025, Centola_2018}. While very simple mathematical models are unable to capture the full complexity of opinion formation processes, they have been successful in advancing their conceptual understanding and contributed to the discovery of new phenomena. For example, analysis of the voter model revealed that a simple opinion imitation process is not sufficient to explain opinion diversity or polarization. Subsequent extensions of the model by majority rules, stubborn agents, social impact or adaptivity of links, demonstrated different avenues leading to minority persistence and dominance \cite{NowakSzamrejLatane1990, SZNAJD2000, Castellano2009, Xie2011, Zschaler2012, Castellano2009b}.

One commonality between the great majority of network-based opinion formation models is that network nodes are used to represent individual agents. If spatial separation of agents is considered, it is usually only reflected indirectly in the pattern of connections between these agents. 
Here, we examine opinion dynamics through an ecological lens at the community level, providing a mathematically tractable framework to investigate the structural factors that shape opinion distributions.
In ecology, network models where nodes are individuals exist as well, but they are complemented by metapopulation and metacommunity models, where nodes represent habitat patches connected by avenues of dispersal \cite{holyoak2005metacommunities, leibold2018metacommunity, gross2020modern}.

Metacommunity models can be represented by reaction-diffusion equations, where the reaction term describes the dynamics on the nodes and the diffusion term describes the dispersal between nodes. This characteristic allows for the formation of spatial patterns through Turing instabilities, which are considered fundamental for explaining spatial phenomena and their functional implications \cite{Turing1952, TuringPatterns2022}.

While originally stated for continuous space, it was shown that Turing instabilities also occur in discrete-space reaction-diffusion systems on networks \cite{SegelLevin1976, Nakao_2010, Fernandes2012, Brechtel2018}.
\begin{figure}[t]
    \includegraphics{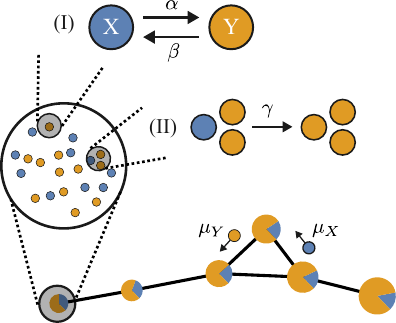}
    \caption{\label{fig:opinion_model_1} Overview of the model. Agents hold one of two opinions, X or Y, and populate nodes of a complex network that are coupled by diffusion. Opinion formation on each node is governed by two processes: (I) spontaneous flipping of opinions and (II) opinion adjustment through the influence of others. Represented at the bottom is the final distribution of opinions after numerical integration of Eq. \ref{eq:network_dynamics} for a six-node network structure. Node size represents the total abundance $X_i+Y_i$, while colors indicate the proportion of opinion X (blue) and Y (orange). Parameters: $\alpha = 0.6$, $\beta=0.8$, $\gamma=0.02$, $\mu_X=15$, $\mu_Y=1$. Initial conditions were chosen such that X is the minority opinion and $X_i(0) + Y_i(0)=15$ for each node $i$.}
\end{figure}
In such systems, it is of particular interest to find the conditions that lead to the formation of spatial patterns and evaluate them qualitatively to understand their functional role within the system.

Master stability functions can be used to analytically disentangle the reciprocal impact of the local dynamics, the network structure and the dispersal strategy on the formation of spatial patterns. Among others, this approach was applied to understand the functional implications of pattern formation in models of cooperation dynamics or linguistic evolution \cite{Fahimipour2022, mimar2021}. Following this work, the term \textit{spatial patterns} here refers to states where the proportion of opinions differs between the network nodes.

In this paper we model opinion dynamics using a meta-community framework. By explicitly incorporating community structure, we analytically derive conditions for pattern formation and show their qualitative implications on the opinion formation process.

Specifically, we develop a network-based model where nodes represent communities populated by agents who choose between two competing opinions and occasionally migrate between different communities. Using a master stability function approach, we disentangle the contributions of local interactions and network structure to the overall dynamics and demonstrate how Turing instabilities can generate spatial patterns that help maintain opinion diversity.

\section{\label{sec:model}Model}
We model the propagation of opinions on a network, where nodes represent communities, such as villages or ecological habitats and links correspond to routes of migration or ecological dispersal between them. 

Agents on each node can hold one of two opinions, which we label X and Y. Each node is initially assigned the same opinion composition. 

The opinion dynamics are governed by two processes (Fig.~\ref{fig:opinion_model_1}). (I) \textit{Spontaneous flipping}, where agents flip to the opposite opinion without interaction. This process introduces an opinion bias when the flipping rates between the opinions are asymmetric. In the following we denote the opinion with the lower or higher flipping rate as the \textit{superior} or \textit{inferior} opinion, respectively, since agents with the superior (i.e., stronger) opinion hold it for longer in the absence of interactions. (II) \textit{Opinion adjustment}, where agents change their opinion due to the normative pressure emanating from others \cite{Latane1996, Xia2011, Larsen1974, Chen2016}, i.e., when they interact with others. Following \cite{Chen2016}, we consider both, interactions between two and three agents. In a double interaction, if two agents with opposing opinions meet, one of them adopts the opinion of the other. In a triple interaction, an agent adopts its opinion if it meets two other agents that hold the opposite opinion. 

The opinion dynamics on each node are approximated by a mean-field model where we assume homogeneous mixing of agents within the communities. 
The state of each node $i$ is represented by two variables $X_i$ and $Y_i$, describing the abundance of agents holding opinion X and Y, respectively. We denote the opinion distribution by the following terminology in the rest of the paper: considering a specific node, the opinion with the lower abundance on that node is called \textit{local minority} opinion. Considering the full network, the opinion that has the lower abundance when aggregated over all nodes is called \textit{global minority} opinion. Conversely, the opinion in majority is called \textit{local or global majority} opinion.

Using the laws of mass action we can translate the dynamics into a set of ordinary differential equations
\begin{equation}
    \label{eq:local_dynamics}
    \begin{split}
        \dot{X}_i &= f(X_i,Y_i) = -\alpha X_i + \beta Y_i + \gamma X_i^2Y_i - \gamma X_i Y_i^2 \, ,\\
        \dot{Y}_i &= g(X_i,Y_i) = - \beta Y_i + \alpha X_i + \gamma X_i Y_i^2 -  \gamma X_i^2Y_i \, ,
    \end{split}
\end{equation}

where $\alpha$ and $\beta$ represent the rates of flipping from opinion X to Y or Y to X, respectively, and $\gamma$ is the rate at which agents change their opinion in triple interactions. The case of binary interactions does not have an impact on the mass action equation and is therefore implicitly included without appearing in Eq. \ref{eq:local_dynamics} (Appendix \ref{appendix_1}).

We now consider a setting where agents occasionally migrate to a different community. Following a common ecological approach, the rate of departure from a node is assumed to be proportional to the number of links of that community. The network structure is described by the adjacency matrix $\textbf{A}$ where $A_{ij}=1$ if node $i$ and node $j$ are coupled by diffusion and $0$ otherwise. Links are bidirectional and lossless. Equation \ref{eq:local_dynamics} is thus extended to
\begin{equation}
    \label{eq:network_dynamics}
    \begin{split}
        \dot{X}_i &= f(X_i,Y_i) - \mu_X k_iX_i + \sum_j{\mu_X \textbf{A}_{ij}X_j} \, , \\
        \dot{Y}_i &= g(X_i,Y_i) - \mu_Y k_iY_i + \sum_j{\mu_Y \textbf{A}_{ij}Y_j},
    \end{split}
\end{equation}
where $k_i$ is the degree of node $i$ and $\mu_X$ and $\mu_Y$ are the diffusion rates of X and Y, respectively.

Numerical integration of Eq. \ref{eq:network_dynamics} shows that under certain conditions the dynamics lead to the formation of spatial patterns where the opinions are distributed heterogeneously across the network (Fig. \ref{fig:opinion_model_1}). 

In the following, we analytically derive the conditions under which spatial patterns occur and show how the specific structure of the model can be used to derive the reciprocal impacts of the network properties and the dynamical process on the opinion formation.

\section{\label{sec:bifurcation}Local Dynamics}
\begin{figure}[h]
    \includegraphics{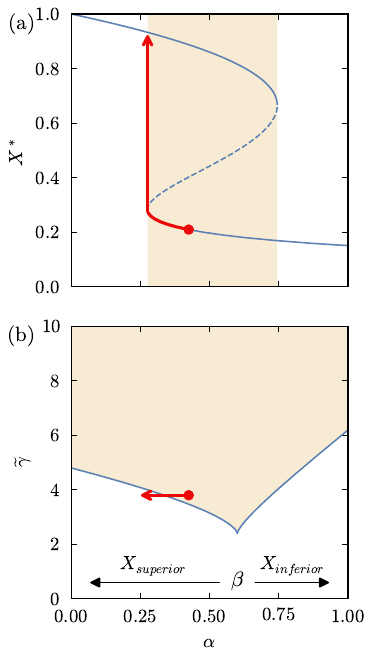}
    \caption{\label{fig:local_bifurcation} (a) One-parameter bifurcation diagram depending on $\alpha$ with $\beta=0.6$ and $\widetilde{\gamma}=0.4$ (Eq. \ref{eq:normalised_dynamics}). Stable and unstable steady states are illustrated by solid and dotted lines, respectively. (b) Two-parameter bifurcation space for $\alpha$ and $\widetilde{\gamma}$ with $\beta=0.6$. Solid blue lines represent saddle-node bifurcation points and meet tangentially in the cusp-point. In the bistable region (orange), X can be the minority opinion (lower branch) even when it is superior according to the flipping rate (i.e., $\alpha < \beta$). Under these conditions a change in $\alpha$ or $\widetilde{\gamma}$ can trigger a catastrophic transition, where the dominant opinion switches abruptly (red arrows).}
\end{figure}

In discrete-space reaction-diffusion systems such as the one described in Eq. \ref{eq:network_dynamics}, diffusion usually leads to homogenization, i.e., the long-term opinion composition on all nodes is identical. In the following we refer to such states where the opinion variables become identical across all nodes as \emph{homogeneous states}. The stability of a homogeneous state can be characterized by analyzing one node in isolation \cite{Brechtel2018}. Therefore, we are interested in the steady states of Eq. \ref{eq:local_dynamics} and their stability.
The dynamics obey the conservation law $X+Y = M$ and can thus be represented in terms of one variable. We can further normalize by $M$ such that $x+y=1$, where $x=X/M$ and $y=Y/M$. By substituting $y=1-x$ we arrive at
\begin{equation}
    \label{eq:normalised_dynamics}
    \dot{x} = -\alpha x + \beta (1-x) + \widetilde{\gamma} x^2(1-x) - \widetilde{\gamma} x (1-x)^2,
\end{equation}
with $\widetilde{\gamma}=\gamma M^2$.

To gain a first intuition of the local dynamics we consider different values of $\alpha$ and $\beta$. For $\alpha$ = $\beta$ = 0, the system has two absorbing states at $x^* = 0$ or $x^* = 1$. A change of $\widetilde{\gamma}$ causes a change in the speed of convergence to the absorbing state, but does not change their location (see also Appendix \ref{appendix_1}).

When $\alpha = \beta > 0$ both opinions coexist. For low values of $\widetilde{\gamma}$, the system is monostable and the two opinions exist in similar proportion ($x^*=0.5$). In contrast, for large values of $\widetilde{\gamma}$ the system is bistable with one stable state at $x^*>0.5$, where $x$ is the local majority opinion and one stable state at $x^*<0.5$, where $x$ is the local minority opinion. The transition between these regimes occurs via a supercritical pitchfork bifurcation where the balanced state ($x^*=0.5$) loses stability, giving rise to two new stable states with high and low proportions of $x$, respectively.

When $\alpha \neq \beta$, the symmetry is broken and we have one superior and one inferior opinion. Depending on $\widetilde{\gamma}$, the system can either be monostable or bistable. In the monostable region, the superior opinion is always in the majority (e.g., $x^*>0.5$ for $\alpha<\beta$), while in the bistable region the superior opinion can be in the majority or minority, depending on the initial conditions.  

The asymmetry leads to a hysteresis loop where the monostable and bistable parameter regions are separated by two saddle-node bifurcations (Fig. \ref{fig:local_bifurcation}a). In the two-parameter space spanned by $\alpha$ and $\widetilde{\gamma}$ the saddle-node bifurcations are located along two lines and separate the monostable from the bistable region (Fig. \ref{fig:local_bifurcation}b). The two lines meet tangentially in a codimension-2 bifurcation (cusp-point), where the stable steady state with $x$ as the majority opinion collides with the stable steady state with $x$ as the minority opinion.

For the opinion formation process these characteristics imply that the superior opinion can still be in the minority state if the opinion adjustment rate is sufficiently high for the system to lie in the bistable region. In this case, where the superior opinion is the minority, a catastrophic transition can be triggered when the opinion adjustment rate is decreased or the flipping rate of the superior opinion is further reduced (red arrows in Fig. \ref{fig:local_bifurcation}).

\begin{figure*}[htpb]
    \includegraphics{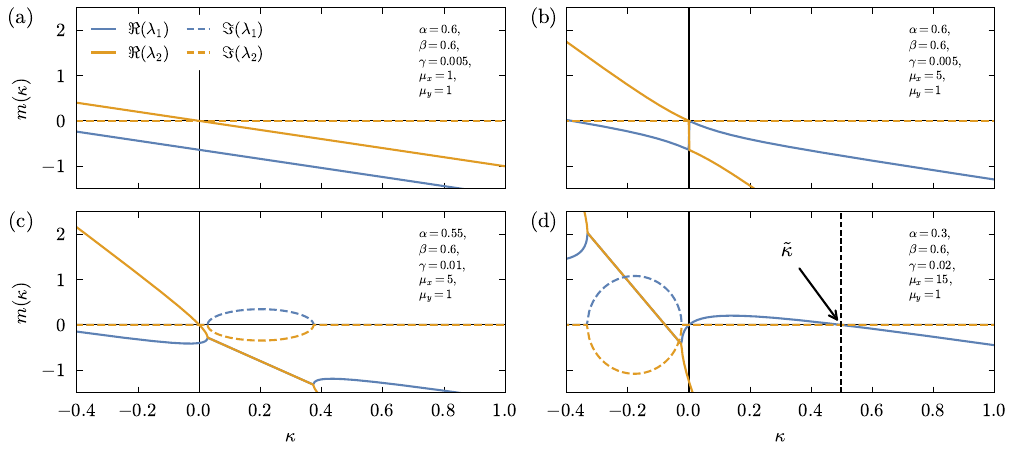}
    \caption{\label{fig:msf} Imaginary and real parts of both eigenvalues obtained from the master stability function depending on the Laplacian eigenvalue $\kappa$ for different parameter settings. (a) The homogeneous state is always stable when $\mu_X=\mu_Y$. (b) For $\alpha=\beta$ the homogeneous state is always stable when the local dynamics (\textbf{P}) are in the monostable region, i.e., $\gamma$ is relatively low. (c) When the non-linearity of the model is strong enough, for a certain range of $\kappa$ the imaginary parts separate, while the real parts merge and split up again, leading to a short increase of $\Re(\lambda_1)$. (d) Due to a further increase of the non-linearity and diffusion rate the same mechanism as in (c) now leads to a positive master stability function in the range $0<\kappa<\tilde{\kappa}$.}
\end{figure*}

\section{\label{sec:bifurcation_msf}Dynamics on Network Structure}

Under certain conditions, the dynamics of the model leads to the formation of patterns where the proportion and abundance of opinions vary between the network nodes (Fig. \ref{fig:opinion_model_1}). We refer to this as the \emph{heterogeneous state}. 

To determine when the system transitions from the homogeneous to the heterogeneous state, we analyze the locally linearized response of the dynamics to small perturbations. This is captured by the eigenvalues $\lambda _i$ of the system's Jacobian matrix \textbf{J} at the homogeneous state. If any $\lambda _i > 0$, the homogeneous state is unstable and spatial patterns emerge. The dependence of these eigenvalues on the network structure can be captured by defining a master stability function 
\begin{equation}
    m(\kappa)= \Lambda_{\max}(\textbf{P}-\kappa \textbf{C})\label{eq:msf},
\end{equation}
where $\Lambda_{\max}$ denotes the eigenvalue with the largest real part, \textbf{P} is the $2 \times 2$ Jacobian matrix of the local dynamics at the steady state (Eq.~\ref{eq:local_dynamics}), \textbf{C} is the coupling matrix with $\mu_X$ and $\mu_Y$ on its diagonal and $\kappa \geq 0$ is an eigenvalue of the network's Laplacian matrix \textbf{L}, encoding its structural properties (Appendix \ref{appendix_msf}). To identify the conditions where the homogeneous state loses stability, we ask whether there are parameter settings of \textbf{P} and \textbf{C} for which $m(\kappa)>0$. 

\begin{figure}[htbp]
    \includegraphics{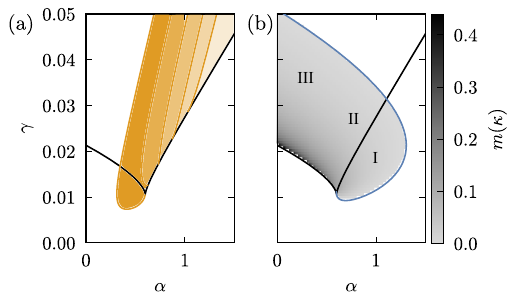}
    \caption{\label{fig:msf_2} Region of pattern formation in the $(\alpha$–$\gamma)$ parameter space ($\beta=0.6$). Black lines represent the saddle-node bifurcation of the local dynamics. The border between the homogeneous and heterogeneous state is given by solid colored lines. (a) Destabilization for different values of $\kappa$, with $\mu_X=1$ and $\mu_Y=15$. The shaded areas show the parameter regions where spatial patterns can emerge. Different shades belong to different values of $\kappa$ from $\kappa=0.05$ (dark orange) to $\kappa=1$ (light orange). (b) Destabilization for $\kappa=0.05$, with $\mu_X=15$ and $\mu_Y=1$. The shading corresponds to the value of the master stability function $m(\kappa)$ for each parameter combination. Roman numbers mark three qualitatively different regions where pattern formation can occur: (I) $\alpha>\beta$ in the monostable region; (II) $\alpha>\beta$ in the bistable region with initialization such that X is in the minority; (III) $\alpha<\beta$ in the bistable region with initialization such that X is in the minority.}
\end{figure}
\begin{figure*}[htbp]
    \includegraphics{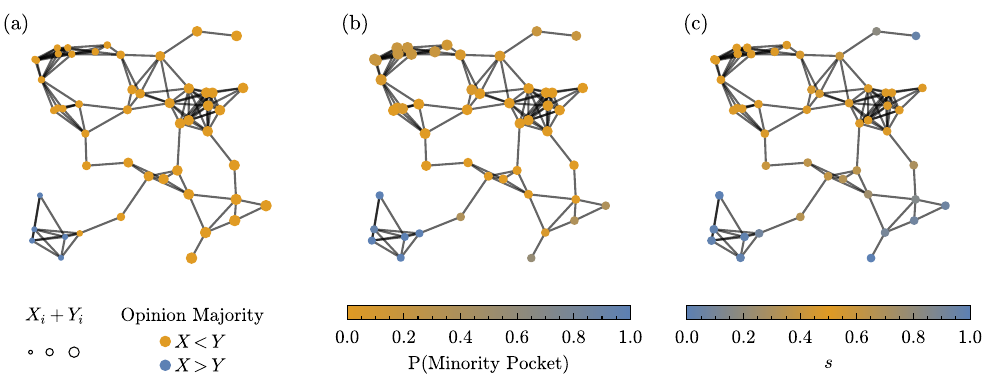}
    \caption{\label{fig:random_geometric} Results from numerical integration and Monte Carlo simulations of the opinion model on a random geometric graph with $50$ nodes and connectivity $0.2$. Local parameters are $\alpha=0.1, \beta=0.6,\gamma=0.03, \mu_x=15, \mu_y=1, M=15$. The system is initialized such that Y is the majority opinion according to the local conditions and $Y_i(0)+X_i(0)=M$ for every node $i$. (a) Distribution of opinions after numerical integration until convergence to the static pattern. Node size represents the total abundance per node, $X_i+Y_i$, and color indicates if a node is dominated by opinion X (blue) or opinion Y (orange). (b) Distribution of opinions after simulation of the model with the Gillespie algorithm ($2\times10^6$ steps, $20$ runs). Node color represents the proportion of runs where on average a node was dominated by opinion X (blue) or Y (orange) over the last $5\times10^4$ steps. Node size represents the average abundance, $X_i+Y_i$, over the same period. (c) Coloring of the nodes based on the linear combination of the unstable Laplacian eigenvectors weighted by the corresponding value of the master stability function as defined in Eq. \ref{eq:msf_agg_measure}.}
\end{figure*}
\begin{figure}[htpb]
    \includegraphics{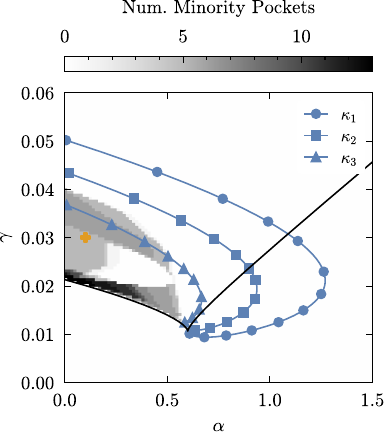}
    \caption{\label{fig:msf_3} Numerical integration of the opinion model on the same random geometric graph as illustrated in Fig. \ref{fig:random_geometric} for different values of $\alpha$ and $\gamma$. Other parameters are $\beta=0.6, \mu_x=15, \mu_y=1, M=15$. The system is initialized such that Y is the majority opinion according to the local conditions and $Y_i(0)+X_i(0)=M$ for every node $i$. Grey shades indicate the number of flipped nodes per $(\alpha,\beta)$-parameter combination after convergence to the static pattern. Black lines represent the saddle node bifurcations of the local dynamics. Blue lines show the border between the homogeneous and heterogeneous state for the three smallest Laplacian eigenvalues that are larger zero. The orange plus indicates the parameter combination used in Fig. \ref{fig:random_geometric}.}
\end{figure}

Instabilities can arise in two fundamental ways, reflected in the sign change of the master stability function: a single eigenvalue crosses the imaginary axis which results in stationary patterns (Turing instability), or a pair of complex-conjugate eigenvalues crosses the imaginary axis, which results in traveling wave patterns (wave instability). 

For a $2 \times 2$ matrix, a pair of complex conjugate eigenvalues can cross the imaginary axis only if two conditions are met: 
\begin{equation}
    \label{eq:msf_wave_1}
    \text{ (I) } \ \lambda_1+\lambda_2=0 ,
    \qquad
    \text{ (II) } \ \lambda_1\lambda_2>0 \ ,
\end{equation}
where $\lambda_1$ and $\lambda_2$ are the eigenvalues of $\mathbf{P}-\kappa \mathbf{C}$. Condition (I) corresponds to
\begin{equation}
    \label{eq:msf_wave_2}
    \text{tr}(\mathbf{P}-\kappa \mathbf{C})=0,
\end{equation} 
which yields
\begin{equation}
    \label{eq:msf_wave_3}
    \kappa=\frac{\text{tr}(\textbf{P})}{\text{tr}(\textbf{C})}=\frac{\text{P}_{11}+\text{P}_{22}}{\mu_X+\mu_Y},
\end{equation}
and cannot be satisfied, because $\kappa \geq 0$, $\mu_X$ and $\mu_Y$ are positive diffusion rates and $\text{tr}(\textbf{P})<0$, as we evaluate the system at a stable steady state. Consequently, wave instabilities do not occur in the system.

Turing instabilities, where one eigenvalue crosses the imaginary axis, occur when
\begin{equation} \label{eq:msf_turing_1a}
    \lambda_1 \lambda_2=0 \, ,
\end{equation}
which corresponds to
\begin{equation} \label{eq:msf_turing_1b}
    \text{det}(\mathbf{P}-\kappa \mathbf{C})=0 \,.
\end{equation}
We can solve Eq. \ref{eq:msf_turing_1b} for $\kappa$ and by considering that \textbf{P} always has a zero eigenvalue due to the conservation property of the local dynamics, we find one trivial solution at $\kappa=0$ and one nontrivial solution at
\begin{equation} 
    \label{eq:msf_turing_2}
    \tilde{\kappa} = \frac{\text{P}_{11}}{\mu_X}+\frac{\text{P}_{22}}{\mu_Y} \, .
\end{equation}
Because $\text{tr}(\textbf{P}) < 0$, a positive $\tilde{\kappa}$ is only possible for unequal diffusion rates $\mu_X \neq \mu_Y$. 

Furthermore, we explore how the local dynamics affect the possibility of spatial patterns by substituting $P_{11}$ and $P_{22}$ into Eq. \ref{eq:msf_turing_2}. This yields
\begin{equation} \label{eq:msf_turing_3}
    0 < \frac{-\gamma Y_*^2 + 2\gamma X_*Y_* - \alpha}{\mu_X}+\frac{-\gamma X_*^2 + 2\gamma X_*Y_* - \beta}{\mu_Y} \, , 
\end{equation}
where $X_*$ and $Y_*$ denote the steady state of the local dynamics. 
We consider the scenario where $\alpha = \beta$ and $\gamma$ is sufficiently small such that the system is monostable with $X_*=Y_*$. 
For this case the condition from Eq. \ref{eq:msf_turing_3} reduces to
\begin{equation} 
    \label{eq:msf_turing_4}
    0 < \frac{\gamma X_*^2 - \alpha}{\mu_X}+\frac{\gamma X_*^2 - \alpha}{\mu_Y} \,,
\end{equation}
i.e., Turing instabilities can only occur when $\gamma>\alpha/X_*^2$. However, this is exactly the onset of the pitchfork bifurcation where the steady state $X_*=Y_*$ loses stability (in Appendix \ref{appendix_lba}, solve Eq. \ref{eq:two_para_bif_2} for $\gamma$ and substitute $X_* = 0.5$). Consequently, for equal flipping rates Turing instabilities can only appear in the bistable region. More generally, Eq. \ref{eq:msf_turing_3} shows that spatial patterns emerge if the opinion that is in the minority has the larger diffusion rate (Fig. \ref{fig:msf}).  

To show how the pattern formation depends on the local dynamics, we plot the heterogeneous region in the $(\alpha,\gamma)$-space in Fig.~\ref{fig:msf_2} (additional information in Appendix \ref{appendix_gba}). 
The onset of the destabilization takes place at the saddle-node bifurcation where the opinion with the higher diffusion rate is in minority, and from there extends into the bistable region.
Depending on the value of $\kappa$, the onset of destabilization moves along the saddle-node bifurcation line towards the cusp-point ($\alpha=\beta$). If the onset of destabilization is close to the cusp-point, spatial patterns also become possible in the monostable parameter region.

We can identify three qualitatively different regions in the $(\alpha,\gamma)$-space where spatial patterns emerge (Fig. \ref{fig:msf_2}b): (I) the minority opinion has the higher diffusion rate in the monostable region; (II) the inferior opinion has the higher diffusion rate in the bistable region and is in minority; and (III) the superior opinion has the higher diffusion rate in the bistable region and is in minority. Generally, the strength of the destabilization, represented by the magnitude of the master stability function, is highest in the vicinity of the saddle node bifurcation at which the minority state of the superior opinion vanishes, i.e., in region (III) of Fig. \ref{fig:msf_2}b.

So far we have shown for which local parameters and diffusion rates the master stability function can become positive and the homogeneous state loses stability. Let us now explore how the emergence of spatial patterns depends on the structure of the community network by considering its Laplacian eigenvalues. 

The Laplacian eigenvalues of a network with $N$ nodes can be ordered such that $\kappa_0 \leq \kappa_1 \dots \leq \kappa_N$, where $\kappa_0 = 0$. The number of zero eigenvalues is the number of disconnected components in the network. For a connected network the second-smallest eigenvalue $\kappa_1 > 0$ is called the algebraic connectivity. 
It is known that $\kappa_1 \geq {4}/Nd$, where $d$ is the diameter of the network \cite{Mohar1991, gross2013}. Consequently, if $Nd \leq 4/\tilde{\kappa}$, the instability is impossible.

Opinion dynamical models commonly use abstract network structures, such as Erdős–Rényi graphs or small-world networks. However, here we draw inspiration from ecological models, where migration between habitats is usually embedded in physical space. Spatially constrained migration is also relevant in the context of opinion dynamics, e.g., when considering movement between cities. A network model that captures the effect of an underlying geographic space is the random geometric graph, which we therefore use in the following for simulations of our model. 

Let us first illustrate the opinion distribution generated by numerical integration of the model on an exemplary random geometric graph with parameter values such that $\max_{i \ge 2} m(\kappa_i) > 0$  (Fig. \ref{fig:random_geometric}a). The destabilization of the homogeneous state leads to a pattern in which the proportion of opinions and their abundance ($X_i+Y_i$) vary significantly between the nodes. At some nodes, the local majority shifts in favor of the global minority opinion, i.e., the opinion that would be in the minority at all nodes under homogeneous conditions. This has important qualitative implications as the pattern formations creates a regime where the minority opinion can be locally preserved, thereby promoting diversity. In the following we call these nodes \textit{minority pockets}.

Monte Carlo simulations of the model show that the emergence of minority pockets are also stable in the finite population case and the opinion distribution is qualitatively comparable to the deterministic case, however, the specific location of the minority pockets varies between different runs of the simulation (Fig. \ref{fig:random_geometric}b). By coloring the nodes based on the linear combination of the unstable Laplacian eigenvectors $\mathbf{v}_i$ weighted by the corresponding value of the master stability function
\begin{equation} 
   \label{eq:msf_agg_measure}
   s=\sum_{i:m(\kappa_{i})>0}m(\kappa_{i})\mathbf{v}_{i} \,,
\end{equation}
we can illustrate how the spectral properties of the graph structurally predispose which nodes are most likely to be minority pockets  (Fig. \ref{fig:random_geometric}c). 

To characterize the parameter regions where minority pockets can emerge, we numerically integrate the model on the same network as in Fig. \ref{fig:random_geometric} using different $(\alpha,\gamma)$-combinations and determine the final number of minority pockets after convergence (Fig. \ref{fig:msf_3}). Note that minority pockets only occur in the bistable parameter region and when the superior opinion is initialized as the minority opinion. Moreover, regions with high abundance of minority pockets can be directly adjacent to regions with no minority pockets at all, resulting in abrupt qualitative transitions of the opinion formation outcome for small parameter changes.

\section{\label{sec:conclusion}Discussion and Conclusion}

In this paper we explored a metapopulation inspired opinion dynamical model on a network, where diffusively coupled nodes represent communities that are undergoing a binary opinion formation process. The specific setup of the model allowed us to analytically separate the effect of the local opinion dynamics that govern each community and the network structure on the opinion formation outcome. We demonstrated that even with a minimalist model, one can draw non-trivial qualitative conclusions on opinion formation processes. By analyzing the master stability function, we derived characteristics of the network structures for which a destabilization of the homogeneous state can manifest. 
Through simulations of the model on a random geometric graph we demonstrated that the pattern formation can lead to outcomes where on some nodes the prevailing opinion switches. 

In a previous study similar results were generated with an agent-based model where individuals populated different communities and could either adopt the predominant local opinion or move to another community, resulting in spatial heterogeneity and opinion clusters under the condition of strong migration and weak adaptation \cite{Hafizoglu2012}. This underlines that the emergence of spatial patterns in opinion dynamics as described by our work is also robust in the finite population case, which we further verified by Monte Carlo simulations of our model.

In conclusion, we demonstrated how the distribution of opinions is linked to community structure. In contrast to conventional opinion dynamical models with individual-based network structure, we analytically derived the conditions under which community structure induces pattern formation that promotes opinion diversity. Although the local interaction dynamics between individuals are simplified, our results highlight the value of complementing individual-based network models in opinion dynamics with community-based approaches.


\section*{Data Availability}
The Python code and Jupyter notebooks required to reproduce the simulation data and figures presented in this study are publicly available on GitHub \cite{mauch2025code}.

\appendix
\setcounter{figure}{0}
\renewcommand{\thefigure}{A\arabic{figure}}

\section{Derivation of ODEs for Local Dynamics}\label{appendix_1}
To derive the ODEs for the local dynamics (Eq. \ref{eq:local_dynamics}) we consider the net turnover of the different processes and use the laws of mass action to translate them into differential equations. For the spontaneous flipping this is represented by $X \xrightarrow[]{\alpha} Y$ and $Y \xrightarrow[]{\beta}X$ resulting in
\begin{equation}
    \begin{split}
        \dot{X}&=-\alpha X + \beta Y \, \\
        \dot{Y}&= \alpha X - \beta Y.
    \end{split}
\end{equation}
Double interactions are given by the reactions $XY \xrightarrow[]{\gamma}2X$ and $XY \xrightarrow[]{\gamma}2Y$, which translate to
\begin{equation}
    \begin{split}
    \dot{X}&=-\gamma XY + \gamma XY= 0 \, , \\
    \dot{Y}&= -\gamma XY + \gamma XY=0 \,
    \end{split}
\end{equation}
and can therefore be ignored. 
Finally triple interactions are represented by the reactions $2X+Y \xrightarrow[]{\gamma} 3X$ and $X+2Y \xrightarrow[]{\gamma} 3Y$, which translate to
\begin{equation}
    \begin{split}
        \dot{X}&= \gamma X^2Y - \gamma XY^2 \, \\
        \dot{X}&= -\gamma X^2Y + \gamma XY^2,
    \end{split}
\end{equation}
and together with the equation for the spontaneous flipping result in Eq. \ref{eq:local_dynamics}.

\begin{figure}[htbp]
    \includegraphics{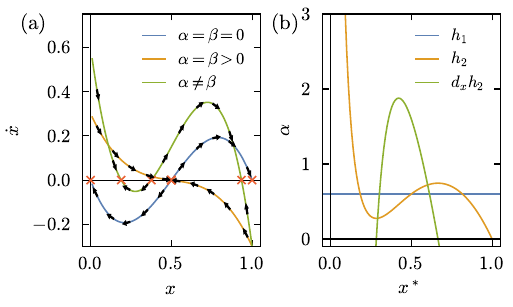}
    \caption{\label{fig:local_bifurcation_A} (a) Phase portrait of Eq. \ref{eq:normalised_dynamics} for different parameter settings (blue: $\alpha=\beta=0, \widetilde{\gamma}=2$, orange: $\alpha=\beta=0.3, \widetilde{\gamma}=1$, green: $\alpha=0.3, \beta=0.6, \widetilde{\gamma}=4.5$). Red crosses mark steady states and black arrows indicate if the flow is directed towards or away from the steady state. (b) Illustration of conditions to derive two-parameter bifurcation diagram from Fig. \ref{fig:local_bifurcation}b.}
\end{figure}

\section{Local Bifurcation Analysis \label{appendix_lba}}

In Fig. \ref{fig:local_bifurcation_A}a we show the phase portrait for different parameter settings of Eq. \ref{eq:normalised_dynamics} to illustrate the insights from Sec. \ref{sec:bifurcation}. 
Additionally, Fig. \ref{fig:local_bifurcation_A}b illustrates a graphical approach to derive the two parameter bifurcation diagram. To analyze the impact of $\alpha$ and $\gamma$ simultaneously we first consider that in the one parameter bifurcation diagram the bistable region is localized by the two saddle-node bifurcations, separating it from the monostable area (Fig. \ref{fig:local_bifurcation}a).
From this, two conditions can be derived that are fulfilled at the bifurcation points and can be solved for $\alpha$ and $\gamma$.
For Condition (I) we define that the system is at a steady state, given by $\dot{X}=0$. 
By solving Condition (I) for $\alpha$ we obtain
\begin{equation} \label{eq:two_para_bif_1}
    \alpha=-2\widetilde{\gamma}X^2+3 \widetilde{\gamma} X+\frac{\beta}{X}-\beta -\widetilde{\gamma},
\end{equation}
where for clarity, we denote the left and right side of the equation by $h_1$ and $h_2$, respectively. The intersections of $h_1$ and $h_2$ are the steady states of the system (compare Fig. \ref{fig:local_bifurcation_A}b).
Additionally, the points where $h_1$ is tangential to $h_2$ are exactly the saddle-node bifurcations that separate the bi- from the monostable region.
This is the case at
\begin{equation} 
    \label{eq:two_para_bif_2}
    \frac{dh_2}{dX} = -4\widetilde{\gamma} X + 3\widetilde{\gamma}  - \frac{\beta}{X^2}=0,
\end{equation}
which we use as Condition (II) to derive the bifurcation diagram.
Condition (II) can now be solved for $\widetilde{\gamma}$ and substituted into Condition (I), such that $\alpha$ and $\widetilde{\gamma}$ only depend on $X$ and $\beta$, which can be solved in parametric form, $\alpha(X,\beta)$, where $\beta$ is fixed and $X$ runs through all possible values in the interval $[0,1]$.

\section{Master Stability Function \label{appendix_msf}}

Consider a reaction-diffusion system on a network with $N$ nodes and $S$ variables per node of the form
\begin{equation} \label{eq:appendix_msf}
    \dot{x}_{si} = f(...) - \mu_s k_i x_{si}+\mu_s \sum_j{A_{ij}x_{sj}},
\end{equation}
where variable ${x_{si}}$ is the $s^{th}$ variable on the $i^{th}$ node, the function $f(...)$ describes the local dynamics on every node, $\mu_s$ is the specific diffusion rate of variable $x_s$, $\mathbf{A}$ is the networks adjacency matrix and $k_i = \sum_j{A_{ij}}$ the degree of node $i$.
Due to the structure of the system, the Jacobian matrix can be represented in a block form
\begin{equation} \label{eq:msf_jacobian}
  \mathbf{J}=\mathbf{I}\otimes \mathbf{P} -\mathbf{L} \otimes \mathbf{C},
\end{equation}
where $\otimes$ denotes the Kronecker product between two matrices, $\mathbf{I}$ is the identity matrix of dimension $S \times S$, $\mathbf{P}$ is the local Jacobian derived from $f(...)$, $\mathbf{L}=\sum_j{A_{ij}}-A$ is the networks Laplacian matrix which can be interpreted as a discrete Laplacian operator on a network \cite{Merris1998}, and $\mathbf{C}$ is the coupling matrix given by
\begin{equation} \label{eq:appendix_msf_coupling}
  \mathbf{C} = \begin{pmatrix}
  \mu_1 & 0 & \cdots & 0 \\
  0 & \mu_2 & \cdots & 0 \\
  \vdots & \vdots & \ddots & \vdots \\
  0 & 0 & \cdots & \mu_S
  \end{pmatrix}.
\end{equation}
It was shown in \cite{Brechtel2018} that due to the specific form of $\mathbf{J}$, its eigenvalues can be computed by
\begin{equation} \label{eq:msf_all}
  \Lambda(\mathbf{J}) = \bigcup_{m=1}^{M} \Lambda(\mathbf{P} - \kappa_m \mathbf{C}),
\end{equation}
where $\kappa_m$ is the $m^{th}$ eigenvalue of the Laplacian matrix $\mathbf{L}$, and $\Lambda()$ is an operator returning the set of eigenvalues of a matrix. This can be rewritten by treating $\kappa$ as an unknown
\begin{equation} \label{eq:msf_final}
  m(\kappa)=\Lambda_{max}(\mathbf{P}-\kappa \mathbf{C}),
\end{equation}
where $\Lambda_{\max}$ returns the eigenvalue with the maximum real part and $m(\kappa)$ is a master stability function through which the bifurcation points that are leading to spatial heterogeneity can be identified as a function of the Laplacian eigenvalue $\kappa$. 
If $m(\kappa) < 0$ for all $\kappa \geq 0$, the homogeneous state is stable and the behavior on all nodes can be determined by analyzing the local dynamics only. Otherwise, if $m(\kappa) > 0$ is valid for at least one $\kappa$, the homogeneous state is unstable and the spatial distribution will be heterogeneous.

\section{Spatial Patterns in Two Parameter Space \label{appendix_gba}}

For the derivation of Fig. \ref{fig:msf_2}, similar to Appendix \ref{appendix_lba}, we define two conditions that are met at the bifurcation point and can be solved for $\alpha$ and $\gamma$, respectively.
We define Condition (I) by expressing Eq. \ref{eq:local_dynamics} only in terms of $X$ with $Y=M-X$ and set $\dot{X}=0$, i.e., the local dynamics are at a steady state. We solve Condition (I) for $\alpha$
\begin{align} \label{aeq:two_para_bif_5}
    \alpha&=\beta M - \beta+\gamma X(M-X)-\gamma (M-X)^2,
\end{align}
where we denote the left and right side of the equation by $h_1$ and $h_2$. Condition (II) is defined by
\begin{equation} \label{aeq:two_para_bif_6}
    \text{det}(\mathbf{P}-\kappa \mathbf{C})=0,
\end{equation}
i.e., the point where the leading eigenvalue returned from the master stability function crosses the imaginary axis and thus the stable steady state(s) defined in Condition (I) are destabilized. We substitute $h_2$ into Condition (I) which we can then solve for $\gamma$ and re-substitute the result into Condition (I). Through this we have derived expressions for $\alpha$ and $\gamma$ that depend only on $X$ and $\beta$, such that the bifurcation curves can be written in parametric form $(\alpha(X,\beta), \gamma(X,\beta))$, where $\beta$ is fixed and $X$ runs through all possible values in the interval $[0,M]$.

\newpage

\providecommand{\noopsort}[1]{}\providecommand{\singleletter}[1]{#1}%

\end{document}